\documentstyle[12pt]{article}
\begin{document}

\baselineskip 0.9cm

\begin {center}{\LARGE { Domain Walls Spacetimes:
Instability of Cosmological Event and Cauchy Horizons }}
\end{center}
\vspace{.6cm}
\centerline {\it Anzhong Wang \footnote{e-mail address: wang@on.br}}

\begin{center}
Department of Astrophysics, Observatorio Nacional, Rua General
Jos\'e Cristino $77, 20921-400$ Rio de Janeiro, RJ, Brazil
\end{center}

\centerline {\it Patricio S. Letelier \footnote{e-mail address:
letelier@ime.unicamp.br} }
\begin{center}
Department of Applied Mathematics-IMECC, Universidade Estadual
de Campinas, $13081-970$ Campinas, SP,  Brazil
\end{center}

\vspace{.6cm}


\baselineskip 0.9cm

The stability of cosmological event and Cauchy horizons of spacetimes
associated with plane symmetric domain walls  are studied. It is found
that  both  horizons are not stable against perturbations of null
fluids and massless scalar fields; they are turned into curvature
singularities. These singularities are light-like and strong in the
 sense that both the tidal forces and distortions acting on test
 particles become unbounded when theses singularities are approached.

\vspace{.3cm}



\newpage

\baselineskip 0.9cm

\noindent {\bf I. INTRODUCTION}

It is the general belief [1], often referred to as non-hair conjecture,
that the external gravitational field of a very massive collapsing body
will finally relax to a black hole field, described by the
three-parameter, the mass, charge and angular momentum, class of the
Kerr-Newmann (KN) solutions [2], although some counterexamples exist
[3]. Regarding to the latter, one of the questions is how much those
examples represent the evolution of a realistic collapsing body?

Assuming that the non-hair conjecture is true, we still are left with
the problems concerning the internal structure of black holes.   The KN
black hole possesses a Cauchy (inner) horizon, beyond which the
predictability of physics, similar to the case of naked singularities
[3], becomes impossible even at the classical level.  However, as first
noticed by Penrose [4], the Cauchy horizon (CH) is a surface of
infinite blueshift, and thus when perturbed by some radiative tails
(these tails are always expected to exist [5]) it will be turned into a
spacetime singularity. This observation has been verified both by
perturbations [6] and by analytic investigations [7, 8]. In particular,
Poisson and Israel [8] found that when two oppositely moving null
fluids are present, the CH in the Reissner-Nordstr\"om (RN) solution is
replaced by a curvature singularity, and that the mass parameter on
this surface becomes unbounded -- the  so-called mass inflation
phenomenon. In view of this enormous mass, the charge and angular
momentum become irrelevant, and then the internal is accurately
described by the Schwarzschild solution. Thus, the CH actually services
as the ultimate everything-proof dam [9], at which the evolution of the
spacetime is forced to stop. As a result, the predictability is
preserved.  For the generic cases, the nature of this singularity,
light-like (null) or space-like, now is still unclear [10], although
according to the strong cosmic censorship conjecture [11] the
space-like is more favorable [12].

Motivated partly by the recent studies of inflationary universe [13],
models including the cosmological constant have been considered [14,
15]. In particular, it is found that in contrary to the KN black hole,
the ones with the cosmological constant have a CH which is stable for
certain choice of the free parameters. Thus, the problem of the
predictability rises again.  However, it should be noted that whenever
the cosmological constant is different from zero, a cosmological event
horizon (CEH) is present [16].

The studies of the internal structure of black holes carried out so far
are mainly restricted to the spacetimes with spherical symmetry [10],
although some attempt to spacetimes with axial symmetry has been
already initiated [17]. However, because of the complexity of the
problems involved, such studies (even in the spherically symmetric
case) are frequently frustrated [10]. Therefore, it would be of
interest to investigate the above mentioned problems in the spacetimes
which are simpler but in which some of those non-trivial properties of
black holes remain. Activity in this direction has been already taken
in  low dimensional spacetimes [18].

In this paper, we shall study stability of the CEH and CH in the usual
(3+1)--dimensional spacetimes with plane symmetry, due to the recent
discovery of non-trivial topology of plane domain wall spacetimes [19 -
21]. In these solutions, CEH's, CH's and event horizons (EH's) are all
present. Since the spacetimes with plane symmetry are easier to handle,
they provide a base on which the above issues can be
studied in some details. It might be argued that spacetimes with plane
symmetry are not realistic, and that they involve infinitely large
masses. In addition, domain walls violate the strong energy condition
[22] (but not the weak and dominant ones). However, as we shall see
below, they do shed some lights on the black hole paradigm.

The rest of the paper is organized as follows: In Sec.II some
properties of the spacetimes with plane symmetry are briefly reviewed.
In Sec.III the instability of the CEH's appearing in a domain wall
spacetime [19] is studied. Following a similar line, the instability of
the CH's of a supersymmetric plane domain wall [20] is investigated in
Sec. IV.  Finally, in Sec. V our main conclusions are presented.

In this paper, the units are chosen such that $8\pi G = 1 = c$, where
$G$ denotes the gravitational constant, and $c$ the speed of light. The
signature of the metric is $+ - - -$.

\vspace{2.cm}

\noindent {\bf II. SPACETIMES WITH PLANE SYMMETRY }

To facilitate our discussions, in this section we shall briefly review
some properties of spacetimes with plane symmetry {\footnote{ Here we
use the definition of plane symmetry originally given by Taub in [23].
Recently, this definition was extended to cover a more general
situation  [24, 25]. Now, the spacetimes defined by Eq.(1) are said to
have planar symmetry.}}. The metric in general can be written as [23]
\begin{equation}
ds^2 = f(dt^2 - dz^2) - e^{- U}(dx^2 + dy^{2}),
\end{equation}
where $f$ and $U$ are functions of $t$ and $z$ only, and the range of
the coordinates is $ -\infty < t, z, x, y < + \infty$. The three
Killing vectors that characterize the plane symmetry are $\partial_{x},
\partial_{y},$ and $y\partial_{x} - x\partial_{y}$. Introducing two
null coordinates $u$ and $v$ via the relations
\begin{equation}
t =  a(u) + b(v), \;\;\;\;\;\; z =  a(u) - b(v), \;\;\;\;\;\;
\end{equation}
where $a(u)$ and $b(v)$ are two arbitrary functions of their indicated
arguments, subject to $a'(u)b'(v)\not= 0 $, where a prime denotes the
ordinary differentiation, Eq.(1) reads
\begin{equation}
ds^2 = 2e^{- M}dudv - e^{- U}(dx^2 + dy^{2}),
\end{equation}
with $e^{-M} \equiv 2a'(u)b'(v)f(t,z)$. The corresponding non-vanishing
components of the Ricci tensor can be found, for example, in [26]. Due
to the symmetry, the Weyl tensor $C_{\mu\nu\lambda\rho}$ has only one
``Coulomb" component, given by [26]
\begin{equation}
\Psi_{2} \equiv - \frac{1}{2} C_{\mu\nu\lambda\rho}
(l^{\mu}n^{\nu}l^{\lambda}n^{\rho} -
l^{\mu}n^{\nu}m^{\lambda}\bar{m}^{\rho} )
= \frac{1}{6} e^{M}(M,_{uv} - U,_{uv}),
\end{equation}
where $(),_{x} \equiv \partial()/\partial x,$ and $l^{\mu}$, $n^{\mu}$,
$m^{\mu}$, and $\bar{m}^{\mu}$, are four null vectors, defined by
$$ l^{\mu} = e^{M/2}\delta^{\mu}_{v}, \;\;\;\;\;\;\;\;\;\;\;
n^{\mu} = e^{M/2}\delta^{\mu}_{u},$$
\begin{equation}
m^{\mu} = \frac{e^{U/2}}{\sqrt{2}}
(\delta^{\mu}_{x} + i \delta^{\mu}_{y}), \;\;\;\; \bar{m}^{\mu} =
\frac{e^{U/2}}{\sqrt{2}}(\delta^{\mu}_{x} - i \delta^{\mu}_{y}).
\end{equation}
Thus, according to the Petrov classifications [27], the spacetimes
described by Eq.(1) or Eq.(3) are either Petrov type $D$ ($\Psi_{2}
\not= 0$) or Petrov type $O$ ($\Psi_{2} = 0$). Note that the spacetimes
of black holes are Petrov type $D$ [1, 27].

On the other hand, one can show that the two null vectors
$\nabla_{\mu}u \;(= e^{M/2}l_{\mu})$ and $\nabla_{\mu}v \;(=
e^{M/2}n_{\mu})$, where $\nabla$ denotes the covariant derivative,
define two null affinely parametrized geodesic congruences [28], and
that the quantities
$$
Q_{l} \equiv - g^{\mu\nu}\nabla_{\mu}\nabla_{\nu}u = e^{M} U,_{v},
$$
\begin{equation}
Q_{n} \equiv - g^{\mu\nu}\nabla_{\mu}\nabla_{\nu}v = e^{M} U,_{u},
\end{equation} represent the rates of contraction of the null geodesic
congruences, defined, respectively, by $\nabla_{\mu}u$ and
$\nabla_{\mu}v$.

\vspace{2.cm}

\noindent {\bf III. INSTABILITY OF COSMOLOGICAL EVENT HORIZONS}

In 1983, Vilenkin [29] found a solution to the Einstein field
equations, which represents a plane domain wall with zero thickness.
The solution is given by
\begin{equation} f = e^{- k|z|}, \;\;\;\;\; U
= k(|z| - t),
\end{equation}
where $k$ is a positive constant. To justify that the above solution
indeed represents a domain wall, Widrow [30] considered the full
coupled Einstein-scalar field equations, and found that when far away
from the center of the wall, the metric for the Einstein-scalar field
equations is indeed well described by Eq.(7), although when near the
center they are coincide only to the first order. Since in this paper
we are mainly concerned with the behavior of the metric at $|z| =
\infty$ (as we shall see below, this limit describes the locations of
the domain walls' CEH), the description of the wall by Eq.(7) is
sufficient for our present purposes.

One of the interesting features of Vilenkin's solution is that at each
of the three spatial directions a CEH exists. The ones in the $x-$ and
$y-$directions are de Sitter-like, and the extensions beyond them are
similar to the four-dimensional counterpart given in [22]. In [29],
Vilenkin provided an extension beyond the horizon in the $z-$direction,
while lately Gibbons [19], among other things, re-considered this
problem and provided another. In this section, we shall first (III.A)
give another extension of Vilenkin's solution along a line similar to
that of Ref. 21. The extension is first made independently in each side
of the wall, and then glued together.  The explicit expressions for
such a gluing are given, which are not expected in the general case
[32]. In the same subsection Vilenkin's extension and the
interpretation of Vilenkin's domain wall as a closed-hypersurface,
bubble [33],  are also considered. In Sec.III.B the instability of the
CEH's is studied.

\vspace{1.cm}

\centerline {\bf A.  Maximal Extenssion of Vilenkin's Solution}

Following Ref. 21 [See also Refs. 20  and 34], let us first make the
following coordinate transformation
\begin{equation}
u = \alpha^{-1} e^{-k(t+z)/2},\;\;\;\; v = - \alpha^{-1} e^{k(t-z)/2},\;\;\;\;
( z \ge 0 ),
\end{equation}
in the region $z \ge 0$, where $\alpha \equiv k/\sqrt{2}$. Then, in
terms of $u$ and $v$, the metric takes the form of Eq.(3) with
\begin{equation}
M = 0, \;\;\;\;\;\;\;\; U = - \ln(\alpha^{2}v^{2}),
\;\;\;\;  ( z \ge 0 ).
\end{equation}
{}From Eq.(11), we find
\begin{equation}
uv = -\alpha^{-2} e^{-kz}, \;\;\;\; \frac {u}{v} = -e^{-kt}, \;\;\;
( z \ge 0 ),
\end{equation}
which shows explicitly the mappings between the $(t,z)-$ and
$(u,v)-$planes.  In particular, the wall $(z=0)$ is mapped to the
hyperbola $uv = -\alpha^{-2}$, while the hypersurface $z = + \infty$ is
mapped to the two axes $ u = 0$ and $v = 0$, across which the
coordinate $t$ becomes space-like and $z$ time-like. As Gibbons pointed
out [19], these two axes actually are the locations of the CEH's [cf.
Fig.1]. From Fig.1(a) we can see that there essentially exist two
walls, each of them locates in one of the two branches of the hyperbola
$uv = -\alpha^{-2}$. These two walls move towards each other at the
beginning with a constant acceleration, and then recede to infinity and
behave like Rindler's particles [35]. On the other hand, one can show
that the extension given by Eq.(9) is the maximal and analytic
extension of Vilenkin's domain wall solution for the region $z \ge 0$.
This can be seen, for example, by transforming it into the Minkowski
spacetime
$$
\bar{T} = \frac{1}{\sqrt{2}}\{\frac{k^{2}}{4}(x^{2} +y^{2})v + (u +
v)\},\;\; \bar{Z} = \frac{1}{\sqrt{2}}\{\frac{k^{2}}{4}(x^{2} +y^{2})v
+ (u - v)\},$$
\begin{equation} \bar{X} = \frac{kvx}{\sqrt{2}}
,\;\;\;\;\; \bar{Y} = \frac{kvy}{\sqrt{2}}.\;\;\;\;\;
\end{equation}
By using the above equation, Gibbons made the extension for the region
where $z \ge 0$. From the above discussions we can see that the only
difference between ours and Gibbons' is that in our case we have
removed the two regions, $III$ and $III\acute{}$, while Gibbons removed
only region $III\acute{}$ and kept region $III$ as a part of the
extended spacetime. As a result, in our extension, there exist two
walls, while in Gibbons' there exists only one wall.

In the region where $z \le 0$, similarly we make the following
coordinate transformation
\begin{equation}
\bar{u} = -\alpha^{-1} e^{k(t+z)/2},\;\;\;\;
\bar{v} = \alpha^{-1} e^{-k(t-z)/2},\;\;\;\;
( z \le 0 ).
\end{equation}
Then, we have
\begin{equation}
\bar{u}\bar{v} = -\alpha^{-2} e^{kz}, \;\;\;\; \frac {\bar{u}}{\bar{v}}
= -e^{kt},
\;\;\; ( z \le 0 ),
\end{equation}
from which the mappings between the $(t,z)-$ and
$(\bar{u},\bar{v})-$planes can be found easily, which is similar to
that for the region where $z \ge 0$ [cf. Fig.1(b)].  In particular, the
center of the wall $(z = 0)$ is mapped to the hyperbola $\bar{u}\bar{v}
= -\alpha^{-2}$, and the hypersurface $z = -\infty$ to the axes
$\bar{u} = 0$ and $\bar{v} = 0$, across which the coordinates $t$ and
$z$ exchange their roles.

Assuming Eqs.(8) and (12) be valid in the neighborhood of the
hypersurface $z = 0$, we can immediately find the matching between the
two extended regions, which is given by
\begin{equation}
\bar{u} = - (\alpha^{2} u)^{-1}, \;\;\;\; \bar{v} = - (\alpha^{2} v)^{-1}.
\end{equation}
In view of the above equation, we can write the metric in the whole
extended spacetime as
\begin{equation}
ds^{2} = \left\{ \begin{array}{c}
2du dv - \alpha^{2}v^{2}(dx^{2} + dy^{2}),\;\;\;\; (z \ge 0 ),\\
       \\
2(\alpha^{2}uv)^{-2}du dv - (\alpha u)^{-2}(dx^{2} + dy^{2}),
\;\;\;\; (z \le 0 ).
 \end{array} \right.
\end{equation}
{}From the above expression we can see  that the metric
coefficients are continuous across the hypersurfaces $uv = - (\alpha
u)^{-2}$ but not their first derivatives, which reflects the fact that
the walls are located on these surfaces.

It should be noted that instead of gluing the hypersurfaces $ab$ and
$a'b'$ together, as indicated in Fig.1, one can glue each of them with
other pieces that are described by Eq. (15).  Such a process can repeat
infinitely times in the transverse direction, so finally we have a
spacetime that has a chain structure [cf. Fig.2]. By this way actually
we have infinite number of walls in the whole spacetime and all of them
are causally disconnected.

In Ref. 29, Vilenkin gave an extension across the hypersurfaces $|z| =
\infty$. Because of the reflection symmetry, it is sufficient to
consider the extension in the region where $z \ge 0$, which is
performed by introducing two new coordinates $T$ and $Z$ by
\begin{equation}
T = t, \;\;\;\;\;\; Z = \frac{2}{k} (1 - e^{-kz/2}),\;\;\; (z \ge 0).
\end{equation}
 The hypersurface $z = \infty$ is mapped to $Z = 2/k$, and the
center of the wall $z = 0$ to $Z = 0$, while the region $z \in [0,
+\infty)$ to $Z \in [0, \frac{2}{k})$. From Eqs.(8) and (16), on the
other hand, we find
\begin{equation}
u = \alpha^{-1} (1 - \frac{k}{2}Z) e^{-kT/2},\;\;\;\;\;
v = - \alpha^{-1} (1 - \frac{k}{2}Z) e^{kT/2}, \;\;\; (z \ge 0).
\end{equation}
The above expressions show that the half of the $(T, Z)-$plane with $Z
\ge 0$ is mapped to the three regions, I\'{}, I, and III in Fig.1.
Similar to the extension of Gibbons [19], Vilenkin took region III as a
part of the extension, too. As a result, in Vilenkin's extension, there
exists only one wall. However, Vilenkin's extension is different from
Gibbons' in that it excludes regions II and II\'{}. Thus, Vilenkin's
extension is not the maximal extension.

On the other hand, from Eq.(11) we find that
\begin{equation}
\bar{R}^{2} - \bar{T}^{2} = -2uv,
\end{equation}
where $\bar{R}^{2} \equiv \bar{X}^{2} + \bar{Y}^{2} + \bar{Z}^{2}$.
{}From the above expression, it was concluded that the wall in the
Minkowski coordinates (11) is not a plane at all, instead, it becomes a
closed-hypersurface, a  bubble [33].  However, following the
considerations given in [21], we argue that the interpretation of the
above solution as representing a plane domain wall is more favorable
than that as representing a bubble. From Eq.(18) we can see that the
coordinate transformations (11) map region I (or I\'{}) in Fig.1 to the
region where $\bar{R} \in ( |\bar{T}|, (\bar{T}^{2} +
2\alpha^{-2})^{1/2} ]$, and the part D $\equiv \{x^{\mu}:  \bar{T}^{2}
\ge uv \ge 0, u \ge 0\},$ of region II ( or D\'{} $ \equiv \{x^{\mu}:
\bar{T}^{2} \ge uv \ge 0, u \le 0\}$ of region II\'{}) to the region
where $\bar{R} \in [ 0, |\bar{T}| ],$ while the part E $\equiv $ II $-$
D ( or E\'{}$ \equiv $ II\'{} $-$ D\'{}) to a region where the
coordinate $\bar{R}$ takes complex values.  Therefore, in order to have
a geodesically complete spacetime, one is forced to include a region
where $\bar{R}$ is complex, which is clearly physically meaningless.

\vspace{1.cm}

\centerline{\bf B. Instability of the Cosmological Event Horizons }

Now let us turn to consider the stability of the CEH's appearing in the
above solution. Because of the reflection symmetry, without loss of
generality, in the following we shall focus our attention only in the
region where $z > 0$.  Then, from Eqs.(6) and (15) we find
\begin{equation}
Q_{l} = - \frac{2}{v},\;\;\;\;\;\;\; Q_{n} = 0.
\end{equation}
Thus, as $v \rightarrow 0^{-}$, we have $Q_{l} \rightarrow + \infty$,
which indicates that the CEH at $v = 0$ is not stable against
perturbations moving along the null geodesics defined by $l^{\mu}$. To
show that this is indeed the case, it is found sufficient to focus our
attention in one of the two diamonds, say, Fig.1(a). In this region,
let us consider the solutions
\begin{eqnarray}
U &=& -\ln [f(u) + \alpha^{2}v^{2}], \nonumber \\
M &=& \frac{1}{2} \ln\left(\frac{f(u) + \alpha^{2}v^{2}}{\alpha^{2}v^{2}}
\right) - g(u) - h(v),
\end{eqnarray}
where $f, g,$ and $h$ are arbitrary functions of their indicated
arguments.   When these functions vanish, the solutions reduce
to the one given by Eq.(15) in the region where $z > 0$. When they are
different from zero, the corresponding EMT in this region is given by
\begin{equation}
T^{\mu\nu} = \rho_{1}l^{\mu}l^{\nu} + \rho_{2}n^{\mu}n^{\nu},
\end{equation}
where
\begin{equation}
\rho_{1} =\frac{(f'g' - f'')e^{-g-h}}
{[\alpha^{2}v^{2}(f + \alpha^{2}v^{2})]^{1/2}},\;\;\;\;\;
\rho_{2} =\frac{2\alpha^{2}vh'(v)e^{-g-h}}{[\alpha^{2}v^{2}(f +
\alpha^{2}v^{2})]^{1/2}}.
\end{equation}
In the following the arbitrary functions $f, g,$ and $h$ will be chosen
such that $\rho_{1}$ and $\rho_{2}$ are non-negative. Then, we can see
that Eq.(21) represents two null dust fluids propagating along the
geodesic congruences defined, respectively, by $l^{\mu}$ and $n^{\mu}$
[cf. Fig.3].  In order to consider the fluids as perturbations, we
further require that $f, g, h$ and their derivatives be small.

Note that when $\rho_{1}\rho_{2}\not= 0$, we can construct two unit vectors,
$u_{\mu}$ and $\chi_{\mu}$, by [36, 28]
$$ u_{\mu} = \left(\frac{\rho_{2}}{4\rho_{1}}\right)^{1/4}\left[ n_{\mu}
    + \left(\frac{\rho_{1}}{\rho_{2}}\right)^{1/2}l_{\mu}\right],$$
\begin{equation}
 \chi_{\mu} = \left(\frac{\rho_{1}}{4\rho_{2}}\right)^{1/4}\left[ l_{\mu}
   - \left(\frac{\rho_{2}}{\rho_{1}}\right)^{1/2}n_{\mu}\right],
\end{equation}
such that
$$
\rho_{1}l^{\mu}l^{\nu} + \rho_{2}n^{\mu}n^{\nu} = \rho (
u^{\mu}u^{\nu} + \chi^{\mu}\chi^{\nu}), $$
\begin{equation}
\rho = (\rho_{1}\rho_{2})^{1/2},\;\;\;
u_{\alpha}u^{\alpha} = - \chi_{\alpha}\chi^{\alpha} = 1, \;\;\;
u_{\alpha}\chi^{\alpha} = 0. \;\;\;
\end{equation}
The above equations show that the sum of two null fluids behaves like
an anisotropic fluid: the pressure of it has only one non-vanishing
component along the $\chi^{\mu}-$direction, and is equal to the energy
density of the fluid.  Moreover, this anisotropic fluid satisfies all
the (weak, dominant and strong) energy conditions [22].

On the other hand, the combination of Eqs.(4) and (20) yields
\begin{equation}
\Psi_{2} = -  \frac{\alpha^{2}vf'(u)e^{-g-h}}{2 [\alpha^{2}v^{2}
(f +\alpha^{2}v^{2})]^{1/2}}.
\end{equation}
Thus, because of the existence of the perturbations the spacetime now
becomes Petrov type $D$. In terms of $\Psi_{2}$ and $\rho_{1,2}$, the
Kretschmann scalar is given by
\begin{equation}
{\cal{R}} \equiv R^{\mu \nu \lambda \delta} R_{\mu \nu \lambda \delta}  =
4(6\Psi^{2}_{2} + \rho_{1}\rho_{2}).
\end{equation}
{}From Eqs.(22) and (25) we find that as $v \rightarrow 0^{-}$, the
Kretschmann scalar diverges as $v^{-1}$. That is, the CEH at the
hypersurface $v = 0$ is not stable against the null fluids and are
turned into a scalar singularity.  The nature of the singularity is
null. It should be noted that the divergence of ${\cal{R}}$ is due to
the mutual focus of the two null fluids, and that the ``Coulomb"
gravitational field $\Psi_{2}$ remains finite, a phenomenon which was
also found in the spherically symmetric case but at the hypersurface of
a CH with a non-vanishing cosmological constant [See the second paper
quoted in  Ref.  15].

As in the case of the CH's [7, 8], the presence of the null fluid
$\rho_{2}$ is not essential to the formation of the singularity as
indicated by Eq.(19), although it affects the nature of the
singularity. This can be seen by the following considerations. Setting
$h(v) = 0$, then Eq.(22) gives $\rho_{2} = 0$.
To see in the latter case a spacetime singularity is still formed on
the CEH, we follow Ref. 7. We first find a freely-falling
frame, and then we calculate the Riemann tensor in this frame. Since
the components of the Riemann tensor represent the tidal forces
experienced by the time-like particles, if any of them becomes
unbounded, we  conclude that a spacetime singularity is formed.  As
assumed above, the functions $f, g$ and their derivatives are very
small, we see that the time-like geodesics can be well approximated by
the ones with $f = g = 0$. For the latter, the time-like geodesics
perpendicular to the $(x, y)-$plane are simply given by the tangent vector
$\lambda^{\mu}_{(0)} \equiv dx^{\mu}/d\tau = E_{+}\delta^{\mu}_{u} +
E_{-}\delta^{\mu}_{v}$, where $E_{\pm} = [E \pm (E^{2} -
1)^{1/2}]/\sqrt{2}$, $\tau$ denotes the proper time of the test
particles and $E$ the energy. From $\lambda^{\mu}_{(0)}$ we can
construct other three linearly independent space-like unit vectors
$\lambda^{\mu}_{(a)} ( a = 1, 2, 3)$ by
\begin{eqnarray}
\lambda^{\mu}_{(0)} &=& E_{+}\delta^{\mu}_{u} +
        E_{-}\delta^{\mu}_{v},\;\;\;
\lambda^{\mu}_{(1)} = E_{+}\delta^{\mu}_{u} -
        E_{-}\delta^{\mu}_{v}, \nonumber \\
\lambda^{\mu}_{(2)} &=& e^{U/2}\delta^{\mu}_{x},\;\;\;\;\;\;\;\;\;
\lambda^{\mu}_{(3)} = e^{U/2}\delta^{\mu}_{y},
\end{eqnarray}
where $U$ is given by Eq.(20). One can show that such defined four
vectors form a freely-falling frame [37]. Computing the Riemann tensor
in this frame, we find that one of the non-vanishing components is
given by
\begin{equation}
R_{\mu \nu \sigma \delta} \lambda^{\mu}_{(0)} \lambda^{\nu}_{(2)}
\lambda^{\sigma}_{(1)} \lambda ^{\delta}_{(2)}
= \frac{E^{2}_{+}(f'g' - f'')}{2(\alpha v)^{2}},
\end{equation}
which diverges as $v^{-2}$ as $ v \rightarrow 0^{-}$. It is interesting
to note that  twice integration of the above component with respect
to the proper time, which gives the distortion of the test particles,
is proportional to $\ln (-v)$ that also diverges as $ v \rightarrow
0^{-}$.  This is in contrary to the case of the CH in the spherically
symmetric spacetimes [9, 10]. On the other hand, from Eq.(26) we can
see that now the Kretschmann scalar is finite at $v  = 0$. As a matter
of fact, one can show that when $h(v) = 0$ the other 13 polynomial
curvature scalars [38] are also finite at $v  = 0$. Thus, the
singularity now becomes a non-scalar one [39, 40], but still very
strong in the sense that both of the tidal forces and distortion acting
on the test particles diverge as the singularity is approached.

In Ref. 41, we have shown that the CEH's are also not stable against a
massless scalar field and are turned into scalar singularities. The
difference is that there the only non-vanishing component $\Psi_{2}$ of
the Weyl tensor also diverges on the CEH's.

\vspace{1.cm}

\noindent {\bf IV. INSTABILITY OF CAUCHY HORIZONS }

In Ref. 20, Cveti\v{c} and co-workers studied spacetimes induced from plane
supersymmetric domain walls interpolating between Minkowski ($M_{4}$) and
anti-de Sitter ($AdS_{4}$) vacua. It was found that the global structure of
the spacetime has a lattice structure quite similar to that of the RN solution
but without singularities. The solution is given by Eq.(1) with
\begin{equation}
f = e^{- U} = \left\{ \begin{array}{c}
1,\;\;\;\; z \rightarrow + \infty,\\
       \\
(\alpha z)^{-2},\;\;\;\; z \rightarrow - \infty,
\end{array} \right.
\end{equation}
where $\alpha$ is defined as $\alpha \equiv (- \Lambda/3)^{1/2}$, and
$\Lambda$ is the cosmological constant which is negative in the present
case. In between these two asymptotic regions, a domain wall is located, and
the metric coefficients smoothly interpolate between the two vacuum regions.
Since we are mainly concerned with the asymptotic behavior of the spacetime,
without loss of generality, we can take the wall as infinitely thin and
located on the hypersurface $z = - \alpha^{-1}$ [20]. Then, the spacetime
is $M_{4}$ for $z > - \alpha^{-1}$ and $AdS_{4}$ for $z < - \alpha^{-1}$. On
the hypersurface there is a domain wall with the EMT taking the form of Eq.(8)
and $\sigma$ is given by $\sigma = 2\alpha$. By studying the motion of the
test particles, it was found [20] that particles leaving from the wall
and moving to the $AdS_{4}$ side reach $z = - \infty$ in a finite proper time.
Thus, to have a geodesically complete spacetime, one needs to extend the
solution beyond $z = - \infty$. After this is done, the spacetime has
a lattice structure, and the hypersurface $z = - \infty$ actually
represents a CH [20]. For the details, we refer the readers to see Ref. 20.
In the following, we shall consider the stability of the CH against null
fluids and massless scalar fields.

\vspace{1.cm}

\centerline{\bf { A. Perturbations of Null Fluids}}

Choosing $a(u) = u/\sqrt{2}$ and $b(v) = v/\sqrt{2}$ in Eq.(2), then from
Eq.(6) we find that
\begin{equation}
Q_{l} = - Q_{n} = - \sqrt{2}\alpha z,
\end{equation}
in the $AdS_{4}$ side.
Thus, as $z \rightarrow - \infty$, we have $Q_{l} \rightarrow + \infty$
and $Q_{n} \rightarrow - \infty$. Then, we would expect that for the
perturbations of a null fluid moving along the null geodesics defined
by $l^{\mu}$, the CH will be turned into spacetime singularity. To
illustrate this point, let us consider the following solutions
\begin{equation}
M = \left\{ \begin{array}{c}
\ln[\alpha^{2}(u-v)^{2}/2] - g(u),\;\;\;\;  z \le - \alpha^{-1},\\
       \\
- g(u),\;\;\;\; z \ge - \alpha^{-1},
\end{array} \right.
\end{equation}
and
\begin{equation}
U = \left\{ \begin{array}{c}
\ln[\alpha^{2}(u-v)^{2}/2] ,\;\;\;\;  z \le - \alpha^{-1},\\
       \\
0,\;\;\;\; z \ge - \alpha^{-1},
\end{array} \right.
\end{equation}
where $g(u)$ is a smooth function.  When it vanishes,
the solutions reduce to the domain wall solution of Cveti\v{c} {\em et
al} [20]. When it is different from zero but very small, the solutions
represent perturbations on the domain wall background. The
corresponding EMT is given by

\begin{equation}
T_{\mu\nu} = \sigma h_{\mu\nu}\delta(z+\alpha^{-1})
 +  (\rho_{1} l_{\mu}l_{\nu} + p g_{\mu\nu})[1 - H(z+\alpha^{-1})],
\end{equation}
where
\begin{eqnarray}
\rho_{1} &=& - \frac{\sqrt{2} g'(u)}{z}e^{- g(u)},
\;\;\; \sigma = 2\alpha e^{- g(u)},\;\;\;
p = \Lambda (e^{- g(u)} - 1), \nonumber\\
h_{\mu\nu} &=& g_{\mu\nu} - \frac{\xi_{\mu}\xi_{\nu}}
{\xi_{\lambda} \xi^{\lambda}},\;\;\;\;\;\;\;\;\;
\xi_{\mu} \equiv \frac{e^{M/2}}{\sqrt{2}} (\delta^{u}_{\mu} -
\delta^{v}_{\mu}),
\end{eqnarray}
$\delta(x)$ denotes the Dirac delta function, and $H(x)$ the Heaviside
function, which is one for $x \ge 0$ and zero for $x < 0$. Provided
that $\rho_{1} \ge 0$, we can see that the solutions given by Eqs.(31)
and (32) represent perturbations of a fluid in the $AdS_{4}$ region,
which is described by the last term in the right-hand side of Eq.(33).
This fluid is Type II in the sense defined in [22]. In the present
case, since $g(u)$ is very small, we have $p \approx 0$. Thus,
practically the fluid is null. As before, this particular class of
single null fluids can not form a scalar singularity, but, as we shall
show below, it does form a non-scalar one.  To see this, we calculate
the components of the Riemann tensor in a freely-falling frame, which
is now given by Eq.(27) but with
\begin{equation}
E_{\pm} = \{ E(\alpha z)^{2} \pm
[(\alpha z)^{2}((E\alpha z)^{2}  - 1)
]^{1/2} \}/\sqrt{2},
\end{equation}
where $E$ is the energy of the test particles. Then, one can show that
one of the non-vanishing components of the Riemann tensor is given by
\begin{equation}
R_{\mu \nu \sigma \delta} \lambda^{\mu}_{(0)} \lambda^{\nu}_{(2)}
\lambda^{\sigma}_{(0)} \lambda ^{\delta}_{(2)} =
\alpha^{2} - \frac{g'(u)E^{2}_{+}}{\sqrt{2} z}.
\end{equation}
On the other hand, one can also show that as $ z \rightarrow - \infty$
we have $ z \approx \tau^{-1}$, where $\tau$ is the proper time of the
test particles. Thus, from Eqs.(35) and (36) we can see that both the
tidal forces and distortions acting on the test particles become
infinite as $ z \rightarrow - \infty$. That is, the CH on $ z = - \infty$
is indeed turned into a spacetime singularity, and the nature of this
singularity, in contrary to the spherically symmetric case [9, 10], is
strong, although $\Psi_{2}$ is still zero, as one can easily show from
Eqs.(4), (31) and (32).

\vspace{1.cm}

\centerline{ \bf {B. Perturbations of Massless Scalar Fields}}

In order to construct perturbations that turn the CH into a scalar
singularity, one way is to consider two oppositely moving null fluids,
another is to consider perturbations of a massless scalar field,
similar to that of Ref. 41. It should be noted that the specific form
of the perturbations, two null fluids, massless scalar fileds, or any
of others, is not important to the formation of a scalar singularity.
What is really important in our analysis is that the perturbations have
to have the two non zero Ricci scalars, $\Phi_{00}$ and $\Phi_{22}$,
where $\Phi_{00} \equiv (R_{\mu\nu} - \frac{1}{4}
g_{\mu\nu}R)l^{\mu}l^{\nu}$ and $\Phi_{22} \equiv (R_{\mu\nu} -
\frac{1}{4} g_{\mu\nu}R)n^{\mu}n^{\nu}$. They represent the mutual
focus between the matter components of the perturbations moving along
the two null geodesic congruences defined by $l^{\mu}$ and $n^{\mu}$
[42], and the Kretschmann scalar is proportional to $\Phi_{00}
\Phi_{22}$ [38].

The perturbations of a massless scalar field on the above domain wall
background can be studied by the following specific solution that is
given by Eq.(1) with
$$
f = \left\{ \begin{array}{c}
1,\;\;\;\; z \ge - \alpha^{-1},\\
       \\
(\alpha z)^{-2},\;\;\;\; z \le - \alpha^{-1},
\end{array} \right.$$

\begin{equation}
U = - \ln(f) - \ln(t_{0} - t).
\end{equation}
The corresponding EMT is given by
\begin{eqnarray}
T_{\mu\nu} = & & 2\alpha (t_{\mu}t_{\nu} - x_{\mu}x_{\nu} -
y_{\mu}y_{\nu})\delta(z + \alpha^{-1})\nonumber\\
&+&  \phi_{,\mu} \phi_{,\nu} - \frac{1}{2}
g_{\mu\nu} \phi_{,\alpha} \phi^{,\alpha},
\end{eqnarray}
where
\begin{equation}
\phi = \frac{1}{\sqrt{2}} \ln(t_{0} - t),\;\;\;
\phi;_{\mu\nu} g^{\mu\nu} = 0,
\end{equation}
and $t_{0}$ is an arbitrary constant. The fact that the particular
solution (39) is singular at $t=t_0$ will play no role in our analysis,
since we are interested in the limit $|z|\rightarrow - \infty$. Eq.(38)
shows that the solution (37) indeed represents a massless scalar field
$\phi$ on the background of the domain wall of Eq.(29). The
corresponding Kretschmann scalar is given by
\begin{equation}
{\cal{R}} \equiv R^{\mu \nu \lambda \delta} R_{\mu \nu \lambda \delta}  =
\left\{ \begin{array}{c}
\frac{\alpha^{2}}{4(t_{0}-t)^{4}}[96(t_{0}-t)^{4}+8(t_{0}-t)^{2}z^{2}+3z^{4}],
\;\;\;  z \le - \alpha^{-1},\\
       \\
\frac{3}{4(t_{0}-t)^{4}},\;\;\; z \ge - \alpha^{-1}.
\end{array} \right.
\end{equation}
Clearly, as $z \rightarrow - \infty, {\cal{R}}$ diverges like $z^{4}$.
Thus, because of the presence of the massless scalar field, the CH is
turned into a spacetime singularity. By considering the components of
the Riemann tensor in a freely-falling frame, one can show that this
singularity is strong.  In fact, we find that one of them is given by
\begin{equation}
R_{\mu \nu \sigma \delta} \lambda^{\mu}_{(0)} \lambda^{\nu}_{(2)}
\lambda^{\sigma}_{(0)} \lambda ^{\delta}_{(2)} =
\alpha^{2}\left[1 + \frac{\alpha^{2} z^{4}}{4(t_{0}-t)^{4}}\right],
\end{equation}
which diverges like $z^{4}$ as $z \rightarrow - \infty$. On the other hand,
we find that as $z \rightarrow - \infty$ we have $z \approx \tau^{-1}$,
where $\tau$, as before, is the proper time of the test particles.

Inserting Eq.(37) into Eq.(4), we have
\begin{equation}
\Psi_{2} = \left\{ \begin{array}{c}
- (\alpha z)^{2}\left[12(t_{0} - t)\right]^{-1},
\;\;\;\; z \le - \alpha^{-1},\\
       \\
\left[12(t_{0} - t)\right]^{-1},\;\;\;\; z \ge - \alpha^{-1}.
\end{array} \right.
\end{equation}
Thus, as $z \rightarrow - \infty$, $\Psi_{2}$ diverges like $z^{2}
\approx \tau^{-2}$. In contrary to the perturbations of a null fluid,
now we have a ``mass inflation phenomenon" [Recall that in the
spherically symmetric case, $\Psi_{2}$ is proportional to the mass
parameter.].  Even though the analysis was carried out with a very
particular solution, due to the arguments presented at the beginning of
this subsection, we believe that the conclusions are valid for a large
class of scalar field perturbations.

Besides the null singularity occurring on $z = - \infty$, there is also
a space-like singularity on $t = t_{0}$. That is, initially the CH is
turned into a null curvature singularity.  However, as the time is
developing, the spacetime collapses. At the moment $t = t_{0}$, the
spacetime collapses into a space-like singularity, and the null one is
finally replaced by the space-like one [43, 10]. This fact depends on
the particular form of (39). We also believe that solutions of the
field equations presenting a similar singular behavior  will produce
spacetimes with similar singular structure.

\vspace{1.cm}

\noindent{\bf{V. CONCLUSIONS}}

In this paper, we have shown that the CH's appearing in the plane
domain wall solution of Cveti\v{c} {\em et al} [20] are not stable
against both a null fluid and a massless scalar field, and are turned
into strong curvature singularities. In the perturbations of a null
fluid, the divergence of the tidal forces and distortions of the test
particles is purely due to the null fluid, and the Weyl tensor vanishes
identically. However, in the perturbations of a massless scalar field,
it is due to both the scalar field and the ``Coulomb" gravitational
field $\Psi_{2}$. Therefore, a  phenomenon  similar to  mass inflation
occurs in the latter case but not in the former.

On the other hand, we have also shown that the CEH's appearing in
Vilenkin's plane domain wall spacetime are not stable against null
fluids and massless scalar fields. They are all turned into strong
spacetime singularities, as both the tidal forces and the distortions
of the test particles diverge as these singularities are approached.
Regarding to this result, a natural question is that: Is the CEH
appearing in the KN-deSitter solutions also unstable?  To have a
definite answer, one way is to consider the perturbations along a line
given in Refs. 7 and 8. Work in this direction will be reported
somewhere else.

\vspace{2.cm}

\noindent {\bf{Acknowledgments}}
 The authors  gratefully acknowledges financial assistance from CNPq.

\newpage

\noindent {\bf FIGURE CAPTIONS}

{\bf Fig.1} The Penrose diagram for the extended Vilenkin domain wall
spacetime. The spatial coordinates $x$ and $y$ are suppressed.
Fig.1(a) represents the extension in the region where $z \ge 0$. In
particular, the region $z \in [0, +\infty)$ is mapped to the regions
$I$ and $I\acute{}$, where $I \equiv \{x^{\mu}: - \alpha^{-2} \le uv <
0, u > 0\}$ and $I\acute{} \equiv \{x^{\mu}: - \alpha^{-2} \le uv < 0,
v > 0\}$ are symmetric with respect to the hypersurface $u = v$ and are
causally disconnected. The time-like coordinate $t$ is past-directed in
region $I$ and future-directed in region $I\acute{}$. Regions $II
(\equiv \{x^{\mu}:  u, v \ge 0\})$  and $II\acute{} \;(\equiv
\{x^{\mu}: u, v \le 0\})$ are two extended regions, while regions
$III(\equiv \{x^{\mu}: uv < - \alpha^{-2}, u > 0\})$ and $III\acute{}
\;(\equiv \{x^{\mu}: uv < - \alpha^{-2}, v > 0\})$ are the regions in
the other side of the wall. Fig.1(b) represents the extension of the
spacetime in the region where $z \le 0$. Because of the reflection
symmetry, it can be easily obtained by replacing $u, v$ by $\bar{u},
\bar{v}$, and the regions $I, I', II, II', III, III'$ by the ones $A,
A', B, B', C, C'$, respectively. To match the two diamonds together,
regions $III, III'{}, C$ and $C'$ have to be removed. The
identifications on the walls are given by Eq.(14). For example, the two
points $P$ and $Q$ are identical, respectively, to $P'$ and $Q'$. The
lines $ad, bc, a'd'$ and $b'c'$ are the locations of the cosmological
event horizons.

\vspace{.4cm}

{\bf Fig.2} Instead of identifying the two surfaces $ab$ and $a'b'$ as
indicated in Fig.1, we can glue the two with other pieces that are
described by Eq.(15). Repeating this process infinite times
in the transverse direction, finally we obtain a spacetime with a chain
structure, in which there exists infinite number of walls. When drawing
this diagram, instead of choosing $\alpha$ in Eq.(8) as $\alpha =
k/\sqrt{2}$, we have set $\alpha = 1$, so the walls are the vertical
lines.

\vspace{.4cm}

{\bf Fig.3} The projection of the spacetime onto the $(u, v)-$plane.
Two null fluids moving toward each other initially in regions $I$ and
$I'{}$, along the null geodesic congruences defined, respectively, by
$l^{\beta}$ and $n^{\beta}$. After they collide on the $2-$surface $u =
0$ and $v = 0$, they form a curvature singularity on the CEH where $v
= 0, u \ge 0$. The nature of the singularity is null.

\end{document}